\documentclass{article}
\usepackage{arxiv}

\usepackage{stmaryrd}
\usepackage{amsfonts}
\usepackage{times}

\usepackage{arxiv}
\usepackage[utf8]{inputenc} 
\usepackage[T1]{fontenc}    
\usepackage{hyperref}       
\usepackage{url}            
\usepackage{booktabs}       
\usepackage{amsfonts}       
\usepackage{nicefrac}       
\usepackage{microtype}      
\usepackage{graphicx}
\usepackage{natbib}
\usepackage{doi}
\usepackage{program}
\usepackage{listings}
\usepackage{color} 
\usepackage{program}

\def\urltilde{\kern -.15em\lower .7ex\hbox{\~{}}\kern .04em}

\title{ProLiVis: Protein-Protein Interaction Literature Visualization System}

\author{ \href{https://orcid.org/0000-0002-9698-4360}{\includegraphics[scale=0.06]{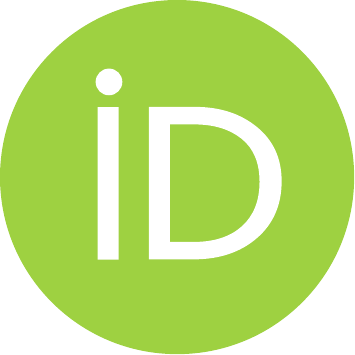}\hspace{1mm}Melih S\"{o}zdinler}$^{1,2}$ \\ 
	$^{1}$ Huawei Turkey R\&D Center, \\
	$^{2}$ Department of Computer Engineering \\
	Bogazi\c{c}i University ,Bebek,Istanbul 34342 Turkey \\
	\texttt{melih.sozdinler@boun.edu.tr}\thanks{Melih S\"{o}zdinler is supported by The Scientific and Technological Research Council of Turkey(TUBITAK)[BIDEB-2211].} 
}

\date{}

\renewcommand{\headeright}{Visualization Tool}

\renewcommand{\shorttitle}{ProLiVis: Protein-Protein Interaction Literature Visualization System}

\hypersetup{
pdftitle={ProLiVis: Protein-Protein Interaction Literature Visualization System},
pdfsubject={q-bio.NC, q-bio.QM},
pdfauthor={Melih Sozdinler},
pdfkeywords={Visualization, PPI Networks, Survey},
}

\begin{document}

\maketitle

\begin{abstract}

\textbf{Summary:} We provide a visualization model that targets the visualization of Protein-Protein Interactions(PPI) and combines it with a super view based on
publications and methods to extract interactions. Although there
are several existing tools, our model considers the existing literature
and is capable to demonstrate all interactions for the selected
organisms. In our model, we propose a three-level visualization concept for the PPI networks with the current state-of-the-art studies based on several organisms. And, we abstract of overall
network based on two perspectives; "experimental method types of each interaction" and "ownership with the publication". We claim that it is more efficient to work on our proposed layout rather than parsing text files or
databases. For that way, we plan to support the existing visuals with complementary outsourced information from the existing knowledge base.

\textbf{Availability:} ProLiVis is available under the MIT License.
Source code is available under Github and binaries are implemented
using {\it OGDF} and cross-platform {\it QT} Framework \newline
\url{https://github.com/melihsozdinler/CenterLayout}

\textbf{Contact:} Melih S\"{o}zdinler, melih.sozdinler@boun.edu.tr
\end{abstract}


\section{Introduction}

Visualization of biological data is a key and important task for research teams. This is because there has been already a data explosion and it is hard to distinguish extract expected know-how from text files without visualization and analysis.In some cases, biological data converges to large networks and
it is more attractive to use some tools to observe hidden outcomes. 
Protein-Protein Interaction(PPI) networks are
one of the kinds that require meaningful visualizations and layout
methods. Unfortunately, these networks may converge
to hundred of thousands of interactions with more than
several thousand nodes. Recently, the statistics of well-known
{\it PPI} database, BioGRID~\cite{citeulike:814974}, has reached 
more than $2,290,000$ raw interactions at the {\it 4.4.203} version. And,
raw interactions was $330,000$ at the {\it 3.2.106} version.
As a result, the visualization of all such interactions is not possible
and yields congested layouts. As a result,
researchers require using new layout models or some abstraction layer for further analysis rather than visualizing all interactions.
Currently, there are several visualization tools aimed to visualize
these networks. These are ProViz~\cite{citeulike:78298}, 
Osprey~\cite{citeulike:670689}, Medusa~\cite{citeulike:474496},
PIVOT~\cite{citeulike:7208034},Robinviz~\cite{1722052}. 
For more information, refer to the review paper of \cite{citeulike:3745109}.
Our motivation is to approach from a different perspective
for the visualization of these networks. Using natural
clustering identities such as {\it Experimental Methods to extract an Interaction}and {\it Published Literature Papers} can support our motivation to do abstraction layer.
With that motivation, we can contain all interactions in one unique layout with
the appropriate abstraction model. This unique layout is also supported
with sub-layouts to increase the efficiency of information extraction. 

\section{Method Overview}

\begin{figure*}[t]
\begin{center}
\hspace*{-.5cm}
\includegraphics[width=17cm]{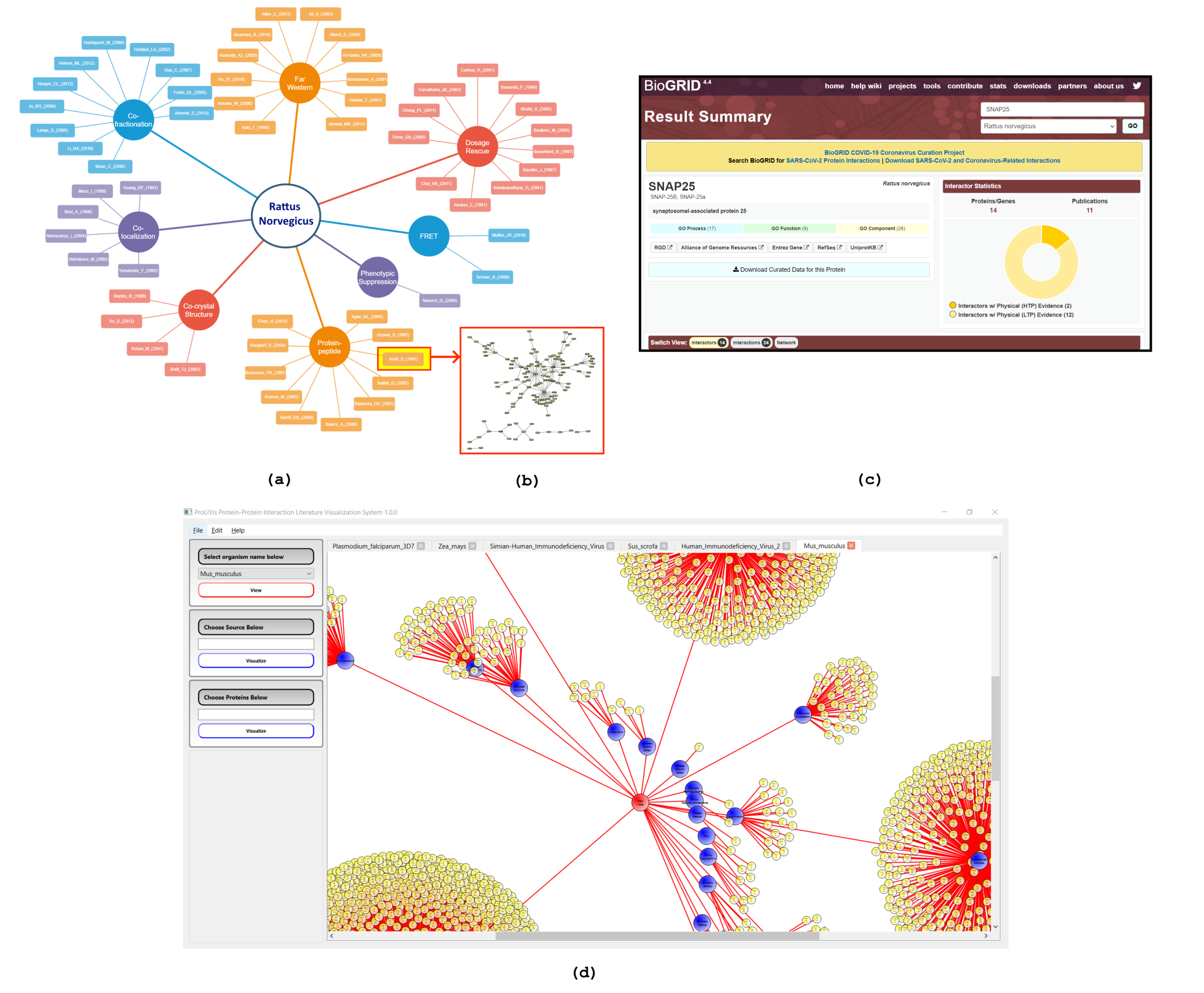} 
\vspace*{.1cm}\\ 
\caption{\sf a) Conceptual First Level Visualization of {\it Rattus Norvegicus}; b) Second Level Visualization or {\it PPI View}; c) Outsourced information tab for the selected protein {\it SNAP25} in (b); d) {\it Mus Musculus} Literature View example from the current ProLiVis Software}
\label{show}
\end{center}
\end{figure*}

Our visualization model contains three levels of hierarchy. 
The first level represents the overall visualization.
The overall visualization contains different types
of nodes and {\it semantic edges}. Nodes differ according 
to two semantic identities;
"Experimental Methodology of an Interaction" as 
{\it type$I$ node}, and "Published Literature Papers and First Author Name" as 
{\it type$II$ node}. Among these, {\it type$I$ node}
identity allows us to filter the whole network with "Experimental Methods". Additionally, {\it type$II$ node} lets us to filter the "Literature" for the specific organisms supported by BioGRID database. 
Using the {\it type$II$ node} node identity,
the contribution of any publication is ready for the visualization. The
amount of contribution can also be validated with the size
of nodes. When the node size grows, this means
its publication source is contributed more.
We also propose the third level. Even, in terms
of publication number, there could be thousands
of {\it type$I$ node} entities.
We limit the first level hierarchy when there
are less contribution of interactions and we
add a new type of node that collects these
nodes into the third layout for each 
{\it type$II$ nodes}.
This allows us to simplify the first level
into one apparent node and to form a third level
for these nodes. The edges in the first and third level
represented as {\it semantic edges} meaning that if a {\it type$I$ node}
is connected {\it type$II$ node}, we infer that
publication under {\it type$I$ node} use the experimental
methodology of {\it type$II$ node}. In addition to all these,
We use the format and data of BioGRID~\cite{citeulike:814974}. 
BioGRID allow us to form three-level layouts for the supported 
organisms.

In Figure~\ref{show} (a), we provide the conceptual visualization of the first level layout of {\it Rattus Norvegicus}. Adjacent nodes like {\it FRET}, {\it Co-Crystal Structure}, {\it Protein-Peptide}
are {\it type$I$ node} are {\it type$II$ node}
and leaf nodes show us the third and they represent literature nodes and we also call {\it type$III$ node}. In Figure~\ref{show} (b),
we provide second-level visualization of {\it type$III$ nodes}. This visualization is currently supported with Force Directed Layout. And, large networks can be considered as future work. In Figure~\ref{show} (c), we select 
{\it SNAP25} protein from Figure~\ref{show} (b) and inside the tool, we will provide access to online databases such as BioGRID, UniProt, and Gene Ontology Amigo. In Figure~\ref{show} (d),
we give the view current development version of ProLiVis and the concept mentioned in Figures~\ref{show} (a,b,c) are expected to be finalized as future work.

There are also several unique properties of our visualization tool.
\begin{itemize}
\item {\it Literature Overview}: We can see all publications
in three-level layouts. The first level and third level
give the literature overview of a specific organism.
For each node, tool-tips and labels can be seen. Each label
has author information and a unique PubMed ID. The layout
is also interactive with zoom-in and zoom-out. 
\item {\it PPI View}: PPI view is the part of all {\it type$I$ node} identities. It is based on {\it Force Directed Layout} of OGDF library~\cite{DBLP:conf/gd/GutwengerJKLMW00}
and allow users to see the whole PPI
the network formed within its publication. Each protein nodes
in PPI View have an option to reach the corresponding
familiar websites using QT Webkit. 
\item {\it Local Computation}: Our main concern is to provide an instant view to users about the current literature of the PPI network. When a user needs to visualize interactions of any publications, that will be obtained from the local indexed database. That way users can continue using the visualization part without the Internet connection. As a future plan, users can also download new DB files when the BioGRID releases a new snapshot. Retrieval of DB files and update procedure is not implemented yet.
\end{itemize}

\section{Conclusion}

ProLiVis differs from other tools with its interactive
level-by-level hierarchy. Similar interactivity is also
proposed in Robinviz~\cite{1722052}. Rather than using theoretical
clustering methodologies, ProLiVis use natural clusters
and propose end-users efficiently without disturbing the
overall layout. For the new enhancements and future work, we plan to extend
the existing first-level hierarchy. We can add new node
types to increase the information extraction such as
publication year. Also, we will write some plugins
to post the specific graphs to the other visualization
tools for further analyses.

BioGRID also offers Cytoscape~\cite{pub.1120775256} plugin(\url{https://wiki.thebiogrid.org/doku.php/biogridplugin2}), however this plugin directly focus on PPI network visualization. There is no combined high-level view for an organism like we offered in Figure~\ref{show} (d).

\section{Acknowledgments}
Melih S\"{o}zdinler is paid by The Scientific and Technological Research Council of Turkey(TUBITAK)[BIDEB-2211]. Additional thanks to my Ph.D. advisors; Can \"{O}zturan, Turkan Haliloglu

{\it Conflict of Interest:} None declared.

\bibliographystyle{unsrtnat}
\bibliography{references} 

\end{document}